\documentclass{mem}
\usepackage{natbib}\usepackage{txfonts}\usepackage{balance}
\usepackage{graphicx}
\usepackage[a4paper,breaklinks,dvipdfm]{hyperref}
\idline{75}{282}
\begin{document}
\def\teff{$T\rm_{eff }$}
\def\kms{$\mathrm {km s}^{-1}$}

\title{
\emph{Fermi}-LAT detection of gamma-ray emission in the vicinity of the star forming regions W43 and Westerlund 2
}

   \subtitle{}

\author{M. Lemoine-Goumard\inst{1, 2}, E. Ferrara\inst{3}, M.-H. Grondin\inst{4}, P. Martin\inst{5} \and M. Renaud\inst{6}}

  \offprints{M. Lemoine-Goumard}

\institute{
Universit\'e Bordeaux 1, CNRS/IN2P3, Centre d'\'Etudes Nucl\'eaires de Bordeaux Gradignan, 33175 Gradignan, France
\and
Funded by contract ERC-StG-259391 from the European Community
\and 
NASA Goddard Space Flight Center, Greenbelt, MD 20771, USA
\and
Institut f\"ur Astronomie und Astrophysik, Universit\"at T\"ubingen, D 72076 T\"ubingen, Germany
\and
UJF-Grenoble 1 / CNRS-INSU, Institut de Planetologie et d'Astrophysique de Grenoble (IPAG) UMR 5274, Grenoble, F-38041, France
\and
Laboratoire Univers et Particules de Montpellier, Universit\'e Montpellier 2, CNRS/IN2P3, Montpellier, France\\
\email{lemoine@cenbg.in2p3.fr}
}

\authorrunning{Lemoine-Goumard M. et al.}

\titlerunning{\emph{Fermi}-LAT observations of W43 and Westerlund 2}

\abstract{
Particle acceleration in massive star forming regions can proceed via a large variety of possible emission scenarios, including high-energy gamma-ray production in the colliding wind zone of the massive Wolf-Rayet binary (here WR 20a and WR 121a), collective wind scenarios, diffusive shock acceleration at the boundaries of wind-blown bubbles in the stellar cluster, and outbreak phenomena from hot stellar winds into the interstellar medium.
In view of the recent Fermi-LAT detection of HESS J1023-575 (in the vicinity of Westerlund 2), we examine another very high energy (VHE) gamma-ray source, HESS J1848-0145 (in the vicinity of W43), possibly associated with a massive star cluster. Considering multi-wavelength data, in particular TeV gamma-rays, we examine the available evidence that the gamma-ray emission coincident with Westerlund 2 and W43 could originate in particles accelerated by the above-mentioned mechanisms in massive star clusters.
\keywords{Gamma rays: general; Galaxies: star clusters: individual: W43, Westerlund 2}
}
\maketitle{}

\section{Introduction}
Young stellar clusters and star forming regions have long been proposed as gamma-ray sources. This relates to the existence of late stages of stellar evolution, which are generally considered favorable for particle acceleration. Different scenarios have been suggested for the gamma-ray production in these environments, considering the contribution of hot and massive Wolf-Rayet (WR) and OB stars and their winds in single, binary, or collective processes, pulsars and their synchrotron nebulae, as well as supernova explosions and their expanding remnants~\citep{reimer, romero}. Their detection and identification using gamma-ray data have already a rather long story, first attempts being made with COS-B~\citep{montmerle} and later with CGRO-EGRET~\citep{romerobenaglia}.\\
Several years after EGRET, the Large Area Telescope (LAT), aboard \emph{Fermi}, offers the opportunity to study faint and extended 
gamma-ray sources such as star forming regions, massive star clusters and pulsar wind nebulae (PWNe). The LAT is an electron-positron pair conversion telescope, 
sensitive to $\gamma$-rays with energies between 30 MeV and 300 GeV, with improved performance (a large effective area, a broad field of view, and a very good angular resolution) compared to its predecessor~\citep{atwood} .\\
The morphological and spectral analyses presented below were performed using two different methods to validate our results: a maximum-likelihood method \citep{mattox} implemented in the \emph{Fermi} SSC science tools as the ``gtlike'' code and "pointlike", a binned likelihood technique that operates by convolving the intrinsic LAT point spread function (PSF) with a spatial model of the candidate source to create a ``Pseudo-PSF" of the source appearance~\citep{kerr}. 
\section{Westerlund 2}
\begin{figure}[t!]
\resizebox{\hsize}{!}{\includegraphics[clip=true]{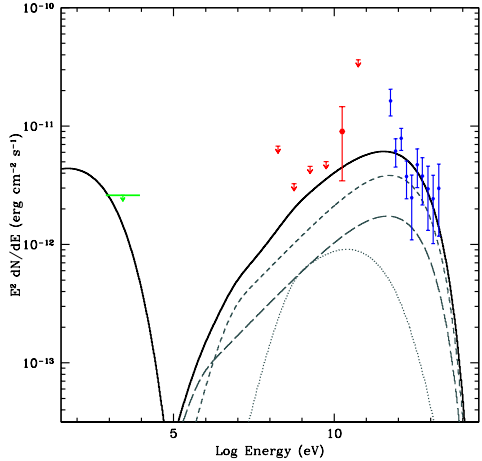}}
\caption{\footnotesize Spectral energy distributions of the off-pulse emission of PSR~J1023$-$5746 \citep{pwncat}. 
The LAT spectral points (red) are obtained using the maximum likelihood 
method "gtlike" in 7 logarithmically-spaced energy bins. A 95~\% C.L. upper limit is computed when the statistical significance is lower than 3~$\sigma$. The blue points represent the H.E.S.S. spectral points~\citep{hessa}. 
The Suzaku upper limit is shown with a green arrow \citep{fujitaetal09}. 
The black line denotes the total synchrotron and Compton emission from the nebula. Thin curves indicate the Compton components from scattering on the CMB (long-dashed), IR (medium-dashed), and stellar (dotted) photons.}
\label{fig1}
\end{figure}

The prominent giant H~II region RCW 49 and its ionizing cluster Westerlund 2 are located towards the outer edge of the Carina arm of our Milky Way at a distance uncertain in the range of values between 2.2 kpc~\citep{brand} and 8 kpc \citep{rauw}. RCW 49 is a luminous massive star formation region, which hosts over two dozen of massive stars as well as two remarkable WR stars (WR 20a and WR 20b), and has been studied extensively over the whole electromagnetic spectrum. This interest was further boosted by the HESS discovery of the bright and extended VHE gamma-ray source HESS J1023$-$575 in the vicinity of Westerlund 2~\citep{hessa}. The extent of the observed gamma-ray emission spatially overlaps with Westerlund~2 and the WR stars WR~20a and WR~20b. Even though the source extension does not support a colliding wind scenario or an emission produced by the Westerlund 2 cluster itself, other scenarii include diffusive shock acceleration in the wind-blown bubble itself and supersonic winds breaking out into the interstellar medium. \\
Recently, \emph{Fermi}-LAT discovered the very young (characteristic age of 4.6 kyr) and energetic (spin-down power of $1.1 \times 10^{37}$~erg~$\rm s^{-1}$) pulsar PSR J1023$-$5746, coincident with the TeV source HESS J1023$-$575~\citep{pablo}. This re-opened the question of the identification of the gamma-ray source and led to re-observations and analyses by HESS~\citep{hessb} and Fermi~\citep{pwncat}. A significant signal in the off-pulse window of PSR~J1023$-$5746 was obtained using Fermi data, with an emission above 10 GeV significant at more than $3\sigma$ level. In addition, the GeV flux connects well with the one derived at TeV energies (as seen in Figure~\ref{fig1}) with a luminosity corresponding to less than 1.5\% of the pulsar spin-down (assuming a distance of 2.4 kpc). This supports highly the scenario in which the gamma-ray source detected by Fermi and HESS would be the PWN powered by PSR~J1023$-$5746. On the other hand, the molecular content of
the region provides sufficient target material to explain the emission through hadronic interaction of cosmic-rays accelerated by a range of feasible mechanisms in the open cluster interacting with molecular clouds. In addition, the lack of X-ray emission from the immediate vicinity of PSR J1023$-$5746~\citep{fujitaetal09} is perplexing given its extremely high spin-down luminosity. More data especially in X-rays and TeV are therefore needed to elucidate the real nature of HESS J1023$-$575.

\section{W43}
\begin{figure}[t!]
\resizebox{\hsize}{!}{\includegraphics[clip=true]{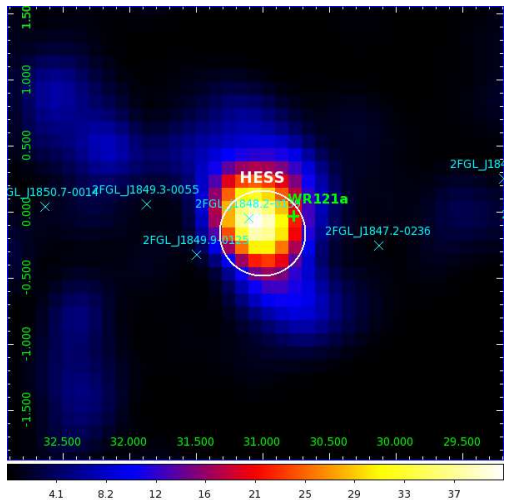}}
\caption{\footnotesize
Test Statistic (TS) map of the region surrounding W43 with side-length $3^{\circ}$, 
above 2~GeV. Each pixel of this image contains the TS value for the assumption 
of a point-source located at the pixel position, Galactic diffuse emission and nearby sources being included in the background model. The position and extent of the HESS source is represented with a white circle. 
The position of WR~121a is indicated with a green cross. Sources extracted from the 2FGL catalog and taken into account in the analysis are marked with a blue cross.
}
\label{fig2}
\end{figure}

\begin{figure}[t!]
\resizebox{\hsize}{!}{\includegraphics[clip=true]{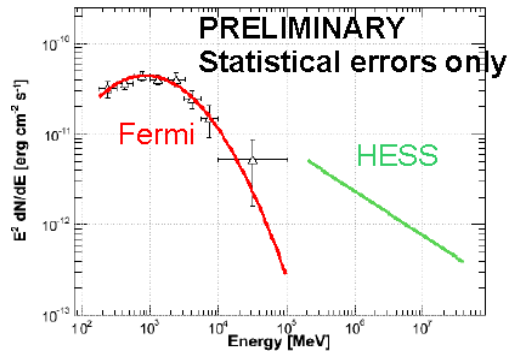}}
\caption{\footnotesize
Spectral energy distribution of the source coincident with HESS J1848$-$018 obtained using pointlike. Only statistical errors are shown. The red solid line presents the best fit obtained assuming a log-parabola. The HESS spectrum~\citep{ryan} is indicated with the green solid line.
}
\label{fig3}
\end{figure}

The W43 star forming region is located along the Scutum-Crux spiral arm tangent at a distance of $\sim 6$~kpc~\citep{nguyen}. The core of W43 harbors a well-known giant H~II region powered by a particularly luminous cluster of WR and OB stars, including WR~121a (a WN7 subtype star), which is characterized by extreme mass loss rates. \cite{motte} established it as a dense region equivalent to a mini-starburst region since it is undergoing a remarkably efficient episode of high-mass star formation. Recent far-infrared to submillimeter data from the Herschel Space Observatory revealed a complex structure of chimneys and filaments, and confirmed its efficiency in forming massive stars~\citep{bally}. A detailed 3D (space-space-velocity) analysis of the molecular and atomic cloud tracers through the region has been carried, revealing that W43 is particularly massive and contains a total mass of ~$7.1 \times 10^6$~ $M_{\sun}$ of molecular gas at a distance of ~6 kpc~\citep{nguyen}.\\
At very-high energies, HESS reported a significant excess ($9\sigma$ pre-trials) with an extended and complex morphology~\citep{ryan}. The TeV source HESS J1848$-$018 is in the direction of, but slightly offset from, the star forming region W43 and there are no energetic pulsars or supernova remnants within $0.5^{\circ}$ which seems to favour an association with the star forming region W43.\\
Trying to unveil the nature of the TeV source, we analyzed 31 months of \emph{Fermi} data collected starting August 4, 2008, and extending until March, 14, 2011. Only gamma-rays in the Pass 7 Source class events were selected, and from this sample we excluded those coming from a zenith angle larger than 100$^{\circ}$ because of the possible contamination from secondary gamma-rays from the Earth atmosphere (Abdo et al., 2009d). Figure~\ref{fig2} shows a TS map of the region in the energy range above 2 GeV: a significant emission coincident with the HESS source and slightly offset from WR~121a is clearly visible. Since the TeV source is significantly extended, we tried to determine the extension of the Fermi source using pointlike: a $3.7 \sigma$ improvement is obtained by using a Gaussian width of ~$0.3^{\circ}$ in comparison to a simple point source. However, the statistics is not large enough to discriminate between one extended source and several point sources. The spectral analysis performed above 100~MeV assuming the best Gaussian model shows that the \emph{Fermi} source is well described by a log-parabola model with spectral parameters 
$\alpha = 2.07 \pm 0.07_{\rm stat}$, $\beta = 0.22 \pm 0.05_{\rm stat}$, $E_b = 1.0$ GeV, and an integral flux above 100~MeV of $(1.77 \pm 0.26_{\rm stat})\times 10^{-7}$ cm$^{-2} \, s^{-1}$:
\begin{equation}
\label{logpar}
\frac{dN}{dE} = N_0 \left( \frac{E}{E_b} \right)^{- (\alpha + \beta \log(E/E_b))}
\end{equation}
The SED presented in Figure~\ref{fig3} is very similar to that obtained on most pulsars detected by \emph{Fermi}-LAT which could indicate that we are contaminated by the emission coming from one energetic radio-faint pulsar (since there are no radio-loud pulsars in this region), as it is the case for Westerlund~2. A blind search is on-going: a pulsed detection in this extended region is a real challenge (the extent of our source is $0.3^{\circ}$): indeed, when the pulsar is off by only 1 arcmin it is already hardly detectable~! As for Westerlund~2, multi-wavelength data would greatly help to reveal the nature of the gamma-ray source, especially in X-rays and in TeV.\\

These two examples indicate that stellar clusters and massive star formation regions might be gamma-ray
sources. However, this identification requires first a discrimination with the Pulsar/PWN (GeV/TeV) scenario which is a real 
challenge in these crowded regions of the Galactic Plane.

\begin{acknowledgements}
The $Fermi$ LAT Collaboration acknowledges support from a number of agencies and institutes for both development and the operation of the LAT as well as scientific data analysis. These include NASA and DOE in the United States, CEA/Irfu and IN2P3/CNRS in France, ASI and INFN in Italy, MEXT, KEK, and JAXA in Japan, and the K.~A.~Wallenberg Foundation, the Swedish Research Council and the National Space Board in Sweden. Additional support from INAF in Italy and CNES in France for science analysis during the operations phase is also gratefully acknowledged.
\end{acknowledgements}

\bibliographystyle{aa}

\end{document}